\documentclass[conference,a4paper]{IEEEtran}
\usepackage{graphics,amsmath,amssymb}
\newtheorem{thm}{Theorem}
\newtheorem{lem}[thm]{Lemma}
\newtheorem{exm}{Example}

\begin{document}

\sloppy

\title{Autocorrelation and Linear Complexity \\of Quaternary Sequences of Period $2p$  Based \\ on Cyclotomic Classes of Order Four}

\author{
  \IEEEauthorblockN{Vladimir Edemskiy and Andrew
Ivanov}
\IEEEauthorblockA{Department of Applied Mathematics and Computer Science\\
     Novgorod State University\\
     Veliky Novgorod, Russia\\
     Email: Vladimir.Edemsky@novsu.ru, dk@live.ru}
 }



\maketitle

\begin{abstract}
 We examine the linear complexity and the autocorrelation of new quaternary
cyclotomic sequences of period  $2p$. The sequences are constructed
via the cyclotomic classes of order four.

\end{abstract}

\section{Introduction}

 The periodic autocorrelation function and the linear complexity
  are important merits for sequence design.
  The autocorrelation measures the amount of
similarity between the sequence $S$ and a shift of $S$ by $w$
positions. The linear complexity ($L$) is defined as the length of
the shortest linear-feedback-shift register that can generate the
sequence \cite{1}. Large linear complexity and  small
autocorrelation for all $w, 1 \leq w \leq N-1$, where $N$  is a
period of a sequence, are desirable features for sequences used in
applications like cryptology and other (see \cite{1, 12, 13}).
Binary and quaternary sequences are the subjects of interest
\cite{11, 10}.

  The use of classical cyclotomic classes and generalized cyclotomic classes
   to construct sequences, which are called classical
cyclotomic sequences and generalized cyclotomic sequences
respectively, is an important method for sequence design \cite{1}.
Legendre sequences are  based on cyclotomic classes of order two.
The properties of Legendre sequences are well known \cite{3}. Kim et
al. \cite{9} defined new quaternary cyclotomic sequences of length
$2p$, where $p$ is an odd prime, and derived the autocorrelation
function of these sequences (see also \cite{4}). Their design is
based on generalized cyclotomic classes of order two.  This approach
was further developed in \cite{8} for sequences of length $2p^m$.

Ding et al. \cite{2} considered the sequences based on cyclotomic
classes of order four and gave several new families of binary
sequences of period $2p$ with optimal three-level autocorrelation.
Now it is interesting to see if it is possible to find quaternary
sequences of length $2p$ based on cyclotomic classes of order four
with desirable properties using the method from \cite{2, 9}.

First, we recall shortly the design of sequences from \cite{2}.  Let
$p=4R+1$ be a prime, where $R$ is a natural number, and let
 $g$ be a primitive root modulo $p$.
 Put, by definition
 \begin{equation*}H_{k} =\{ g ^{k+4t}(\bmod p) ,\, \, t=0,1,\dots,R-1\},
  \ \ k=0,1,2,3.
  \end{equation*}
  Then $H_{k}$ are called cyclotomic classes of order $4$
~\cite{1}.

By the Chinese Remainder Theorem, $\mathbb Z_{2p} \cong \mathbb Z_2
\times \mathbb Z_p$   relatively to isomorphism $f(w) =\left(w_1 ,
w_2 \ \right)$, where $w_1=w \bmod~2,$ $w_2=w \bmod~p$. Here and
hereafter $a\bmod~p$ denotes the least nonnegative integer that is
congruent to $a$ modulo $p$. For the sake of convenience let us
denote $f^{-1} \left(\{k\} \times H_l\right ), k=0, 1, l=0, 1, 2,3 $
as $H_{k,l}$. Then, we have partitions
\begin{equation*}
\mathbb Z_{2p}^*=\bigcup _{l=0}^{3} H_{1, l} \ \ \mbox{ and  } \ \
2\mathbb Z_{2p}^*=\bigcup _{l=0}^{3} H_{0, l}.
\end{equation*}
Let $C_k=H_{0, j_k} \cup H_{1, l_k}$, $k=0,1,2,3,$ where $j_k\neq
j_m,  l_k\neq  l_m $ for $k\neq m$.  Then
\begin{equation*}
C_0 \cup C_1  \cup C_2 \cup C_3 =\mathbb Z_{2p} \setminus \{0,p\}.
\end{equation*} The quaternary
sequence $\{s(t)\}$ is defined as
\begin{equation}
\label{eq1} s(t) =\begin{cases}
 k, & \mbox{ if  }  t\bmod 2p \in C_k, \\
 0, & \mbox{ if  }  t\bmod 2p=0, \\
 2,& \mbox{ if  }  t\bmod 2p=p. \\
 \end{cases}
\end{equation}
\begin{exm} Let $p=13, g=2.$ Then $H_0=\{1,3,9\},$

\noindent $ H_1=\{2,5,6\},$ $H_2=\{4,10,12\},\  H_3=\{7,8,11\}$ and
\begin{multline*}
\ \ \ \ \ \ \ \ \ \ H_{0,0}=\{14,16,22\}, \  H_{0,1}=\{2,6,18\},\\
H_{0,2}=\{4,10,12\},\ H_{0,3}=\{8,20,24\};\\
H_{1,0}=\{1,3,9\},  \ H_{1,1}=\{5,15,19\}, \ \ \ \\
H_{1,2}=\{17,23,25\}, \ H_{1,3}=\{7,11,21\}. \ \ \ \ \ \ \ \ \
\end{multline*}

 If $(j_0, j_1, j_2, j_3)=(0, 1, 2, 3)$ and
$(l_0, l_1, l_2, l_3)=(1, 2, 3, 0)$ then
\begin{equation*}
 s(t) =\begin{cases}
0, & {\rm{if}} \ \  t\bmod 2p \in \{0, 5,14,15,16,19,22\} , \\
1, & {\rm{if}} \ \  t\bmod 2p \in \{2,6,17,18,23,25\}, \\
2, & {\rm{if}} \ \  t\bmod 2p \in \{4,7,10,11,12,13,21\} , \\
3, & {\rm{if}} \ \ t\bmod 2p \in \{1,3,8,9,20,24\}.
  \end{cases}
\end{equation*}
\end{exm}
 This is a generalization of the construction proposed in  \cite{2} and \cite{9}
  to build binary and quaternary sequences, respectively.
   In the next sections we derive  the periodic autocorrelation function of  $s(t)$;
      we also determine the linear complexity of $s(t)$ over the finite field of four elements  and over the finite ring of order four.

\section{Autocorrelation}
\label{sec:1}
 The autocorrelation quaternary $2p$-periodic sequence $s(t)$
is a complex-valued function defined by
\begin{equation*}R(w) = \sum_{n=0}^{2p-1}
i^{s(n)-s(n+w)}, \end{equation*}
 where $i= \sqrt {-1}$ is an imaginary unit.

Let $v(t) = i^{s(t)}$. Then, the periodic autocorrelation function
at shift $w$ of $\{s(t)\}$ is given by
\begin{equation}
\label{eq2} R(w) = \sum _{n=0}^{2p-1} v(n) v^*(n+ w),
\end{equation}
where $v^*(t)$ is the complex conjugate of $v(t)$.

Let $D, E$ be a subset of the ring $\mathbb{Z}_{2p}.$ By definition,
a difference function $ d_w(F, E)$ can by written as
\begin{equation*}d_w(F, E)=|F \cap (E+w)|,\end{equation*}
where $E+w$ denotes the set $\{w+e: e \in E\}$ and $"+"$ denotes an
addition modulo $2p$.

It is well known that if  $f(t)$ and $e(t)$ are the characteristic
sequences of $F$ and $E$, i.e.,
\begin{equation*}
f(t)=\begin{cases}
1, & {\rm if}\ t \bmod 2p\in F,  \\
0, & {\rm otherwise.}
\end{cases}
\end{equation*}
and
\begin{equation*} e(t)=\begin{cases}
1, & {\rm if}\ t \bmod 2p \in E,  \\
0, & {\rm otherwise.}
\end{cases}
\end{equation*}
then
\begin{equation}
\label{eq3} \sum _{j=0}^{2p-1} f(t)e(t+w)=d_w(F,E).
\end{equation}
Consequently, by (\ref{eq1}) - (\ref{eq3}) for real $\bigl(\rm{Re}
R(w)\bigr)$ and imaginary  $\bigl ( \rm{Im} R(w) \bigr )$ parts of
the autocorrelation function $R(w)$ we have the following equations:
\begin{multline}
\label{eq4} \text{Re} R(w)= d_w(C_0 \cup \{0 \}, C_0\cup \{0
\})+d_w(C_1, C_1) \\ + d_w(C_2\cup \{p \}, C_2\cup \{p \}) +d_w(C_3,
C_3) \\ -d_w(C_0\cup \{0 \}, C_2\cup \{p\})-d_w(C_2\cup \{p \},
C_0\cup \{0 \})\\ -d_w(C_1, C_3)- d_w(C_3, C_1),\end{multline} and
\begin{multline}
\label{eq5} \text{Im} R(S,w)= d_w(C_1, C_0\cup \{0 \})+d_w(C_3,
C_2\cup \{p \}) \\ +d_w(C_0\cup \{0 \}, C_3) +d_w(C_2\cup \{p \},
C_1) \\- d_w(C_1, C_2\cup \{p \}) -d_w(C_3, C_0\cup \{0 \}) \\ -
d_w(C_0\cup \{0 \}, C_1)-d_w(C_2\cup \{p \}, C_3).\end{multline}

So, in order to obtain $R(w)$  it is sufficient to find difference
functions listed in  (\ref{eq4}) - (\ref{eq5}).

Suppose that $F_0, F_1, E_0, E_1$ are subsets of $\mathbb{Z}_p.$
\begin{lem}
\label{l1} Let $F=\{0\} \times F_0 \cup  \{1\} \times F_1$, and
$E=\{0\} \times E_0 \cup  \{1\} \times E_1$, and $w=(w_0, w_1) \in
\mathbb{Z}_2 \times \mathbb{Z}_p.$ Then
\begin{equation*}d_w(F,E)=\begin{cases}
|F_0\cap E_0|+|F_1\cap E_1|,  \ \    {\rm if}\ w=0,   \\
|F_0\cap (E_0+w_1)|+|F_1\cap (E_1+w_1)|, & \\  \ \ \ \ \ \ \ \  \ \  \ \ \ \ \ \ \ \  \ \ \ \ \ \ \ \ \  \ \  {\rm if}\ w_0=0, w_1 \neq 0, \\
|F_0\cap (E_1+w_1)|+|F_1\cap (E_0+w_1)|, & \\ \ \ \ \ \ \ \ \ \  \ \
\ \ \ \ \ \ \ \  \ \ \ \ \ \ \ \ \   \ {\rm if}\ w_0=1, w_1 \neq
0, \\
|F_0\cap E_1|+|F_1\cap E_0|, \ \    {\rm if} \   w_0=1, w_1 =0.
\end{cases}\end{equation*}
\end{lem}
\begin{IEEEproof} For the case $F=E$ Lemma \ref{l1} was proved in \cite{2}.  When $F \neq E$ it is proved
similarly. \end{IEEEproof}

 To derive
difference functions we will need cyclotomic numbers. Recall that
the cyclotomic numbers of order 4 in this case are defined as
\cite{7}
$$(i,j)=|(H_i+1) \cap H_j|.$$ Note that every prime $p\equiv
1(\text{mod}~4)$ has a quadratic partition $p=x^2+4y^2$, where $x, y
\in \mathbb{Z}$ and $x\equiv 1(\bmod~4).$ It is well known how to
express cyclotomic numbers over the values of $p, x, y$ \cite{7}.
Here $y$ is two-valued, depending on the choice of the primitive
root $g$ employed to define the cyclotomic classes.
\begin{lem}
\label{l2} If $u \in H_h$, $h=0, 1, 2, 3$, then

(i)  $|H_j\cap (H_l+u)|=(h-l, j-l)$,

(ii) $|H_j\cap \{u\}|=\begin{cases}
1, & {\rm if}\ h=j,  \\
0, & {\rm otherwise.}
\end{cases}$

(iii) $|\{0\}\cap (H_j+u)|= \\
\begin{cases}
1, \ \  {{\rm if}} \ h=j  \text{  and  } p\equiv 1~(\bmod~ 8) \\
\ \ \ \ \ \ \ \ \ \ \ {\text{ or }} \ \ h\equiv ( j+2)(\bmod~ 4) \ \ { \text{and}  } \ \  p\equiv 5~(\bmod~8),  \\
0, \ \ {\text {otherwise.}} \ \ \ \ \ \ \  \ \  \ \ \
\end{cases}$
\end{lem}
\begin{IEEEproof} Notice that
\begin{equation*}|H_j\cap (H_l+u)|= |u^{-1}H_j\cap
(u^{-1}H_l+1)|.\end{equation*} Since $|H_j\cap (H_l+u)|=(l-h, j-h)$
 and $(l-h, j-h)= (h-l, j-l)$ \cite{7} then the first statement is proved.  The second
statement of Lemma \ref{l2} is obvious.

Subsequently, since $-1=g^{(p-1)/2}$, then  $-1 \in H_0$ for  $
p\equiv 1~(\bmod~8)$ and $-1 \in H_2$ for $ p\equiv 5~(\bmod~8)$.
Hence, if $u \in H_h$, then $-u \in H_h$ for  $ p\equiv 1~(\bmod~8)$
and

\noindent $-u \in H_{(h+2)(\bmod~4)}$ for $ p\equiv 5~(\bmod~8)$.
From this follows the third statement of Lemma \ref{l2}.
\end{IEEEproof}

By Lemmas  \ref{l1}- \ref{l2}  and formulas \eqref{eq4}-\eqref{eq5}
we can derive the autocorrelation of $\{s(t)\}$ defined by
\eqref{eq1}. Note that the set of values  $R(w), w\neq 0$,  do not
vary under the complex conjugation of the sequence and under the
cyclic shift modulo four of cyclotomic class numbers in \eqref{eq1}.
Therefore, it is enough to consider two cases, namely when the
vector $(j_0, j_1, j_2, j_3)$ equals $(0, 1, 2, 3)$ or $(0, 2, 1,
3)$. Consider examples of finding the autocorrelation function for
every case.

Let $(j_0, j_1, j_2, j_3)=(0, 1, 2, 3)$ and $(l_0, l_1, l_2,
l_3)=(1, 2, 3, 0)$, i.e., the quaternary sequence $\{s(t)\}$ defined
as
\begin{equation}
\label{eq6} s(t) =\begin{cases}
0, & {\rm{if}} \ \  t\bmod 2p \in \{0\} \cup H_{0,0} \cup H_{1,1}, \\
1, & {\rm{if}} \ \  t\bmod 2p \in H_{0,1} \cup H_{1,2}, \\
2, & {\rm{if}} \ \  t\bmod 2p \in \{p\} \cup H_{0,2} \cup H_{1,3}, \\
3, & {\rm{if}} \ \ t\bmod 2p \in H_{0,3} \cup H_{1,0}.
  \end{cases}
\end{equation}
\begin{thm}
\label{t1} Let the quaternary sequence $\{s(t)\}$ be defined by
(\ref{eq6}). Then

 (i)  $R(w) \in \{-2 \pm 2i, \pm 2i, -2 \}, w=1, \dots,2p-1$ if

 \noindent  $ p\equiv 5~(\bmod~ 8)$,

 (ii)  $R(w) \in \{-4, \pm 2, 0 \}, w=1, \dots,2p-1$ if

 \noindent  $ p\equiv 1~(\bmod
 8).$
 \end{thm}
\begin{IEEEproof} As before, let $w=(w_0, w_1)$. Consider several cases.

1) Let $w_0=0, w_1 \neq 0$. If $w_1 \in H_h, h=0, 1 , 2 , 3$ then by
(\ref{eq4}), Lemma \ref{l1} and Lemma \ref{l2} we have
\begin{equation*}
\text{Re} R(w)=2 \sum _{k=0}^3 \bigl ( (h-k,0)-(h-k,2) \bigr
)+\delta, \end{equation*} where
\begin{equation*}
\delta=\begin{cases} \ \ 0,& \rm{if} \ \  p\equiv  5(\bmod~8),\\
\ \ 2,& {\rm{if}} \ \  p\equiv 1(\bmod~8) \ \ \text{and} \ \ h=0, 3,\\
-2,& {\rm{if}} \ \ p\equiv 1(\bmod~8)\ \ \text{and} \ \ h=1, 2.
\end{cases}
\end{equation*}
It is shown \cite{7} that \begin{equation*} \sum _{k=0}^3
(h-k,0)-(h-k,2) = \sum _{j=0}^3
 (j,0)-\sum _{j=0}^3(j,2)=-1,\end{equation*} hence, if $w_0=0, w_1 \neq 0$ and $w_1 \in H_h$ then
\begin{equation*}
\text{Re} R(w)=\begin{cases} -2,& \rm{if} \ \   p\equiv  5(\bmod~8),\\
\ \ 0,& {\rm{if}} \ \  p\equiv 1(\bmod~8) \ \ \text{and} \ \ h=0, 3,\\
-4,& {\rm{if}} \ \ p\equiv 1(\bmod~8)\ \ \text{and} \ \ h=1, 2.
\end{cases}
\end{equation*}
Similarly we obtain
\begin{equation*}
\text{Im} R(w)=\begin{cases} \ \ 2i,& {\rm{if}} \ \  p\equiv 5(\bmod~8)\ \ \text{and} \ \ h=0, 1,\\
-2i,& {\rm{if}} \ \  p\equiv 5(\bmod~8) \ \ \text{and} \ \ h=2, 3,\\
\ \ 0,& {\rm{if}} \ \ p\equiv 1(\bmod~8).
\end{cases}
\end{equation*}
 2) Let $w_0=1, w_1 \neq 0$.  Here, by Lemmas \ref{l1}  and \ref{l2}
\begin{equation*}R(w)=\begin{cases} \ \  2, & {\rm{if}} \ \  p\equiv 5(\bmod~8)\\
-2, & {\rm{if}} \ \  p\equiv 1(\bmod~8).
\end{cases}\end{equation*}

3) Let $w_0=1, w_1 \neq 0$.  As before, it can easily be checked
that
\begin{equation*}
 R(w)=\begin{cases} \ \ 2i, \ \ {\rm{if}} \ \  p\equiv 5(\bmod~8)\ \ \text{and} \ \ h=0, 3 \ \ \ \ \ \ \ \ \ \\
 \ \ \ \ \ \ \ \ \ \ \ \ \ \ \ \text{or} \ \  p\equiv 1(\bmod~8)\ \ \text{and} \ \ h=0,  1, \\
-2i, \ \ {\rm{if}} \ \  p\equiv 5(\bmod~8) \ \ \text{and} \ \ h=1, 2,\\
\ \ \ \ \ \ \ \ \ \ | \ \ \  \ \text{or} \ \  p\equiv 1(\bmod~8)\ \
\text{and} \ \ h=2,
 3.
\end{cases}
\end{equation*}
Theorem \ref{t1} is proved.
\end{IEEEproof}

 Note that if we take $s(0)=s(p)=0$ then under the conditions of Theorem
\ref{t1} we obtain  $|R(w)|=2, w=1, \dots, 2p-1.$

Take another one case. Let $(j_0, j_1, j_2, j_3)=(0, 2, 1, 3)$ and
$(l_0, l_1, l_2, l_3)=(2, 0, 3, 1)$, i.e.,
 the quaternary sequence $\{s(t)\}$ is defined as
\begin{equation}
\label{eq7} s(t) =\begin{cases}
0, & {\rm{if}} \ \  t\bmod 2p \in \{0\} \cup H_{0,0} \cup H_{1,2}, \\
1, & {\rm{if}} \ \  t\bmod 2p \in H_{0,2} \cup H_{1,0}, \\
2, & {\rm{if}} \ \  t\bmod 2p \in \{p\} \cup H_{0,1} \cup H_{1,3}, \\
3, & {\rm{if}} \ \ t\bmod 2p \in H_{0,3} \cup H_{1,1},
  \end{cases}
\end{equation}
then we can similarly derive the following lemma.
\begin{lem}
\label{l3} Let the quaternary sequence $\{s(t)\}$ be defined by
(\ref{eq7}) and $p=x^2+4y^2, x\equiv 1(\bmod~4).$ Then

 (i)  $\max_{w \neq 0}|R(w)|=\max|-2 \pm 2y \pm 2i| $ if $ p\equiv 5~(\bmod
 8)$,

 (ii) $\max_{w \neq 0}|R(w)|=\max|-4 \pm 2y| $ if $ p\equiv 1~(\bmod
 8).$
 \end{lem}

 The autocorrelation of new sequences is better than that of
the quaternary sequences proposed by Kim et al.

Under the conditions of Lemma  \ref{l3}  the autocorrelation
properties of $\{s(t)\}$ are worse than that under the conditions of
Theorem  \ref{t1}.

 Moreover, we can show that  if the sequence is constructed by \eqref{eq1}
 then its autocorrelation can't be better than that under the conditions of Theorem \ref{t1}.

Further, we derive the linear complexity of $\{s(t)\}$. On the one
hand, it is possible to derive the linear complexity of quaternary
sequences over the finite ring $\mathbb Z_4$. An alternative
approach is Gray-mapping  the quaternary sequences and obtain the
sequences defined over $\mathbb F_4$ (the finite field of 4
elements). Generally, these two ways lead to different values for
the linear complexity because arithmetics of $\mathbb F_4$ and
$\mathbb Z_4$ differ, see for example \cite{6}.

We  explore the linear complexity of $\{s(t)\}$ for both
alternatives.

\section{The Linear Complexity of Quaternary Sequences over the
Finite Field of Order 4}

 We can convert quaternary sequences into the
sequences of elements belonging to the finite field of order four by
using Gray map. In this section we demonstrate that as a result of
Gray mapping of the sequences with good autocorrelation properties
from the Section \ref{sec:1} we obtain the sequences with high
linear complexity over the finite field of order four.

 Let
$\mathbb{F}_4=\{0, 1, \mu, \mu+1 \}$ be a finite field of four
elements, and let $\varphi (a)$ be a Gray map defined by $\varphi
(0)=(0,\ 0)$, $\varphi(1)=(0,\ 1)$, $\varphi(2) =(1,\ 1)$,
$\varphi(3)=(1,\ 0)$. Suppose $\{s(t)\}$ is constructed by
\eqref{eq6}. Consider $\mathbb{F}_{4} $ as  a vector space over
$\mathbb{F}_{2} $ with basis $\mu ,\; 1$.  Denote Gray-mapping of
$\{s(t)\}$   as $\{u(t)\}$ , i.e.,
\begin{equation}
\label{eq8} u(t) =\begin{cases}
0, & {\rm{if}} \ \  t\bmod 2p \in \{0\} \cup H_{0,0} \cup H_{1,1}, \\
1, & {\rm{if}} \ \  t\bmod 2p \in H_{0,1} \cup H_{1,2}, \\
\mu+1, & {\rm{if}} \ \  t\bmod 2p \in \{p\} \cup H_{0,2} \cup H_{1,3}, \\
\mu, & {\rm{if}} \ \ t\bmod 2p \in H_{0,3} \cup H_{1,0}.
  \end{cases}
  \end{equation}
It is well known (see \cite{1})  that the minimal polynomial and the
linear complexity of sequence can be derived as follows:
\begin{multline}
\label{eq9} L=N-\deg \bigl (\gcd \left(x^{N} -1,U(x)\right) \bigr ),
\\  m(x)=\bigl (x^{N} -1 \bigr )/\gcd \left(x^{N}
-1,U(x)\right), \end{multline} here $U(x)=\sum_{t=0}^{N-1} u(t) x^t$
is the generating polynomial of $u(t)$.

 For $N=2p$ we can write (\ref{eq9})
in the form
\begin{equation} \label{eq10} L=2p-\deg \left( \gcd \bigl ((x^{p} -1)^{2} ,U(x)\bigr )\right ) \end{equation}
  Let $\alpha $ denote a primitive root of unity of
order  $p$ in the extension of the field $\mathbb{F}_4$.  From
\eqref{eq10} it follows that to derive the linear complexity and the
minimal polynomial of  $\{u(t)\}$ it is sufficient to determine the
number of roots and multiple roots of  $U(x)$  in the set $\{ \alpha
^{v} ,v=0,1,...,p -1\} $.

  Now we introduce auxiliary polynomials. Let  $U_{4} (x)=\sum _{i\in H_{0} } x^{i} $
   and $U_{2} (x)=\sum _{i\in H_{0} \cup H_{2} } x^{i} $.  The properties of   $U_{2} (x)$ and  $U_{4} (x)$
   were examined in \cite{3}  and \cite{5}, respectively.

\begin{lem}
\label{l4} If $v=0,\dots,p-1$, then

$\sum _{v\in H_{0,\, k} } \alpha ^{v} =\sum _{v\in H_{1,k} } \alpha
^{v} =U_{4} (\alpha ^{g ^{k} } )$.
\end{lem}

\begin{IEEEproof} By definition
\begin{equation*}
\{a \bmod~p | a \in H_{k,l} \}=H_l\end{equation*} for all $ k=0,1;
l=0, 1, 2, 3.$ Since $g^lH_0=H_l$, then $\sum_{j \in H_{k,l}}
\alpha^j=\sum_{j \in H_0} \alpha^{g^k j}$. Lemma \ref{l4} follows
from the last equality.
\end{IEEEproof}

 By Lemma  \ref{l4}  and by the definitions
of the auxiliary polynomials we have
\begin{equation} \label{eq11}
U_{4} (\alpha ^{v} )+U_{4} (\alpha ^{vg^{2} } )=U_{2} (\alpha ^{v}
)\end{equation}
\begin{thm}  \label{t2} Let the sequence $\{u(t)\}$
 be constructed by
(\ref{eq8}). Then

1.  $L=2p$  and $m(x)=x^{2p} -1$ if $p\equiv 1(\bmod~8)$,

2.  $L=(3p+1)/2$  and $m(x)=(x^{2p} -1)/H(x)$,  here $H(x)=\prod
_{i\in H_{1} \bigcup H_{3} }(x-\alpha ^{i} )$ if $p\equiv 5(\bmod~
8).$  \end{thm}
\begin{IEEEproof} By definition, the generating polynomial of the sequence can
be written as:
\begin{multline*} U(x)=\sum _{j\in H_{0,1} \cup H_{1,2} } x^{j} +(\mu +1)\sum _{j\in H_{0,2} \cup H_{1,3} }
x^{j} \\
 +\mu \sum _{j\in H_{0,3} \cup H_{1,0} } x^{j} +(\mu
+1)x^{p} .\end{multline*}
 The number of the
elements in each of the three sums is even, then $U(1)=\mu +1$.

Further, by  Lemma \ref{l4} we have
\begin{multline*}U(\alpha ^{v} )=\mu +1+U_{4} (\alpha ^{g v} )+U_{4} (\alpha
^{g ^{2} v} )\\+(\mu +1)\left(U_{4} (\alpha ^{g ^{2} v} )+U_{4}
(\alpha ^{g ^{3} v} )\right)+\mu \left(U_{4} (\alpha ^{g^{3} v}
)+S_{4} (\alpha ^{v} )\right),\end{multline*}  or to put it
differently,
\begin{multline*} U(\alpha ^{v} )=\mu +1+U_{4} (\alpha ^{g v} )+U_{4} (\alpha ^{g^{3} v} )\\+
\mu \left(U_{4} (\alpha ^{g ^{2} v} )+U_{4} (\alpha ^{v}
)\right).\end{multline*}
 Now, by \eqref{eq11} we obtain
\begin{equation} \label{eq12} U(\alpha ^{v} )=U_{2} (\alpha ^{\theta v} )+\mu U_{2} (\alpha ^{v} )+\mu +1 .\end{equation}
Taking into account that  $p\equiv 1(\bmod~4)$, let us examine two
cases:

1.  Let  $p\equiv 1(\bmod~8)$, then it was shown in \cite{3} that
with an appropriate choice of $\alpha $ we have
\begin{equation*}
U_{2} (\alpha ^{v} ) =\begin{cases}
1, & {\rm{if}} \ \  v \in  H_0 \cup H_2, \\
0, & {\rm{if}} \ \  v \in  H_1 \cup H_3.
  \end{cases}
\end{equation*}
 Therefore, $U(\alpha ^{v} )\in \{ 1,\; \mu \} $, i.e., $U(\alpha ^{v} )\neq 0$  when $v=0,1,...,p-1$, and the statement of
 Theorem \ref{t2}
follows from \eqref{eq9}  and \eqref{eq10}.

2. Let $p\equiv 5(\bmod\, 8)$, then as shown in \cite{3},
\begin{equation*}
U_{2} (\alpha ^{v} ) =\begin{cases}
\mu, & {\rm{if}} \ \  v \in  H_0 \cup H_2, \\
\mu+1, & {\rm{if}} \ \  v \in  H_1 \cup H_3.
  \end{cases}
\end{equation*}
 Therefore,  if $v\in H_{0} \cup H_{2}
$  then by \eqref{eq12} $U(\alpha ^{v} )=\mu +1$; if \textbf{$v\in
H_{1} \cup H_{3} $} then $U(\alpha ^{v} )=0$.

Now we determine the multiplicity of the roots of the form  $\alpha
^{v} $ of the polynomial $U(x)$. The derivative of $U(x)$ is
\begin{multline*} U^{'} (x)=\sum _{j\in H_{1,2} } x^{j-1} +(\mu +1)\sum _{j\in H_{1,3} }
x^{j-1}\\
+\mu \sum _{j\in H_{1,0} } x^{j-1} +(\mu +1)x^{p-1}.\end{multline*}
  Now,
\begin{multline} \label{eq13} U^{'} (\alpha ^{v} )=\alpha ^{-v} \bigl(\mu +1+U_{4} (\alpha ^{g ^{2}
v} )\\+(\mu +1)U_{4} (\alpha ^{g^{3} v} )+\mu U_{4} (\alpha ^{v} )
\bigr ).\end{multline} If $p\equiv 5(\bmod\, 8)$  then $2\in H_{1}
\cup H_{3} $. Thus, as shown in  \cite{4}, if $v\in H_{k} $,  then
$U_{4} (\alpha ^{v} )=\varsigma ^{2^{k} } $  or $U_{4} (\alpha ^{v}
)=\varsigma ^{2^{4-k} } $,  here $\varsigma $ satisfies the
conditions
 $\varsigma ^{8} +\varsigma
^{4} +\varsigma ^{2} +\varsigma =1$ è and $\varsigma ^{4} +\varsigma
=\mu $  in the field $\mathbb{F}_{4} $.

 So, from \eqref{eq13}  we get
$U^{'} (\alpha ^{v} )\neq 0$  for $v\in H_{1} \cup H_{3} $. Hence,
we see that all the roots of $U(x)$ of the form $\alpha ^{v} $ are
simple and the statement of Theorem \ref{t2} for $p\equiv 5(\bmod\,
8)$ follows from \eqref{eq9}  and \eqref{eq10}.
\end{IEEEproof}
If we apply Gray mapping to the sequence defined by \eqref{eq7}, we
obtain
\begin{equation}
\label{eq14}b(t) =\begin{cases}
0, & {\rm{if}} \ \  t\bmod 2p \in \{0\} \cup H_{0,0} \cup H_{1,2}, \\
1, & {\rm{if}} \ \  t\bmod 2p \in H_{0,2} \cup H_{1,0}, \\
\mu+1, & {\rm{if}} \ \  t\bmod 2p \in \{p\} \cup H_{0,1} \cup H_{1,3}, \\
\mu, & {\rm{if}} \ \ t\bmod~2p \in H_{0,3} \cup H_{1,1}.
  \end{cases}
\end{equation}

\begin{lem}
\label{l5}  Let the sequence $\{b(t)\}$
 be constructed by
\eqref{eq14}. Then $L=2p$  and $m(x)=x^{2p} -1.$ \end{lem}

Lemma \ref{l5} is proved in the same way as Theorem \ref{t2}.

Let us consider the other sequences of period $2p$ obtained by
\eqref{eq1}. We can derive their linear complexity by applying
formulas for $U_4(\alpha ^{v})$ from \cite{5}.

\section{The Linear Complexity of Quaternary Sequences over the Ring
of Order 4}

 A polynomial $ C(x) = 1 + c_1x + ... + c_mx^m$, $C(x)
\in \mathbb{Z}_4 [x]$ is called an associated connection polynomial
of a periodic sequence $\{s(t)\}$ over $\mathbb Z_4$, if the
coefficients $c_1,c_2,...,c_m$ satisfy $s(t)=-c_1s(t-1)-c_2s(t-2)
-...- c_ms(t-m),\ \forall t\ge m$. The linear complexity of a
periodic sequence $\{s(t)\}$ over $\mathbb{Z}_4$ is equal to
\begin{multline*}L={\rm {min}}\{ \deg \emph{C(x)}| \ \
C(x) \ \text {is an associated} \\ \text{connection polynomial of} \
\ $\{s(t)\}$ \}.\end{multline*}

  We
know that $C(x)$ is an associated connection polynomial of
$\{s(t)\}$ if and only if
\begin{equation} \label{eq15} S(x)C(x)\equiv 0~\left
(\mbox{mod}(x^{2p}-1)\right ),
\end{equation}
where $S(x) = s(0) + s(1)x + ... + s(2p-1)x^{2p-1}$ \cite{15}.

 Let $R=GF(2^{2r},2^2)$ be
Galois ring of characteristic 4, where  $r$ is the order of 2 modulo
$p$ \cite{14}. The group of invertible elements $R^*=R\setminus 2R$
of the ring $R$ contains the cyclic subgroup of order $2^r-1$
\cite{14}. Then, in  $R^*$  there must exist an element $\beta$ of
order $p$. Let $\gamma=3\beta$, then the order of $\gamma$  is equal
$2p$ and $\gamma^p=-1$.

The maximal ideal of the ring $R$ is $2R$ \cite{14}. Here we have
the natural epimorphism of the rings  $R$ and $\overline{R}=R/2R$.
Let $\overline{r}$ denote the image of the element $r \in R$ under
this epimorphism.

\begin{lem} \label{l6} Let the sequence $\{s(t)\}$
 be defined by \eqref{eq6}. Then
 $S(\overline{\gamma}^v) =1$ if $v=1,\dots,2p-1,  v \neq p.$
\end{lem}
\begin{IEEEproof} By definition of  $\gamma$ it follows that
$\overline{\gamma}^p= \overline{\gamma}$ and $\overline{\gamma} \neq
0, 1$. Then $\sum_{j=1}^{p-1} \overline{\gamma}^j=1$ in the ring
$\overline{R}$.

 By \eqref{eq6} we have
 \begin{multline*} S(x)=\sum _{j\in H_{0,1} \cup H_{1,2} } x^{j} +2\sum _{j\in H_{0,2} \cup H_{1,3} }
x^{j}\\ +3 \sum _{j\in H_{0,3} \cup H_{1,0} } x^{j} +2x^{p}
.\end{multline*} In the same way as in Lemma \ref{l4} we obtain
\begin{multline*} S(\overline{\gamma}^v)=\sum _{j\in H_1}\overline{\gamma}^{jv} +\sum _{j\in H_2}\overline{\gamma}^{jv}
+\sum _{j\in H_3}\overline{\gamma}^{jv} +\sum _{j\in
H_0}\overline{\gamma}^{jv} \\ =
\sum_{l=1}^{p-1}\overline{\gamma}^{l}=1.\end{multline*}
\end{IEEEproof}

Before proceeding to the main results of the section, we note that
in the ring $R$ the number of  polynomial's roots can be greater
than its degree. For example, if
\begin{equation*}P(x)=2(x^p-1)/(x-1),
\end{equation*} then
 $P(\gamma^v)=0$ for $v=1,\dots, 2p-1, v\neq p$, but at the same time  $P(x)$ is not divisible by the product
$(x^{2p}-1)/(x^2-1)$. But, if $a$ and $b$ are the roots of the
polynomial $P(x)$ and $a-b \in R^*$ then $P(x)$  is divisible by
$(x-a)(x-b)$.

  By the definition of  $\gamma$ we have an expansion $(x^p-1)/(x-
1)=\prod_{i=1}^{p-1}(x- \gamma^{2i})$. Then $p
=\prod_{i=1}^{p-1}(1-\gamma^i)(1+\gamma^i)$ . So, $\gamma^j-\gamma^l
\in R^*$  when $j, l=0,...,p-1, j\neq l$.
 \begin{thm}
 \label{t3}
  Let  $\{s(t)\}$ be defined by (\ref{eq6}), then
 $L=2p.$
 \end{thm}
\begin{IEEEproof} Let  $L < 2p$,
then there exists an associated connection polynomial $C(x)$  with
degree less than $2p$ and
  \begin{equation*}S(x)C(x)\equiv 0~\left
(\mbox{mod} (x^{2p}-1) \right )\end{equation*} by (\ref{eq15}).
According to Lemma \ref{l6}   we can write: $C(\alpha^v)=0$ for
$v=1,\dots, 2p-1, v\neq p$. Then  $C(x)$ is divisible by $\left
(x^p-1 \right)/(x-1)$, i.e., $C(x)=Q(x)\left (x^p-1 \right)/(x-1)$,
$Q(x) \in \mathbb{Z}_4 [x]$ and $2Q(x)\neq 0$. Further, by
definition $S(1)=2$, consequently $C(1) \in \{0,2\}$ and $Q(1)\in
\{0,2\}$, hence $Q(x)=(x-1)F(x)+q, q \in \{0,2\}$ or
\begin{equation*}C(x)=\left (x^p-1 \right)F(x)+q\left (x^p-1 \right)/(x-1).\end{equation*}
Then, we obtain   $2F(\alpha^v)=0$  for $v =1,3,\dots,2p-1$,
therefore $2F(x)$ is divisible by $x^p+1$. Since the degree of
$F(x)$ is less than $p$, we get a contradiction.

Consequently, $L=2p.$ \end{IEEEproof}

\begin{lem} \label{l7} Let the sequence $\{s(t)\}$
 be defined by \eqref{eq7}. Then
 $L=2p.$
\end{lem}
Lemma \ref{l7} we prove in the same way as Theorem \ref{t3}.

\section{Conclusion}

We examined new quaternary sequences constructed on the cyclotomic
classes of order four. We showed that they have high linear
complexity and satisfactory autocorrelation. The linear complexity
is derived over the finite field of order four and over the ring of
four elements.


\end{document}